\documentclass[aps,twocolumn,showpacs]{revtex4}
\usepackage{graphicx}
\usepackage{amssymb}

\begin{document}

\title{Explaining a changeover from normal to super diffusion in time-dependent billiards}

\author{Matheus Hansen$^1$, David Ciro$^{2}$, Iber\^e L. 
Caldas$^{1}$ and Edson D.\ Leonel$^{3}$}
\affiliation{$^1$ Instituto de F\'isica - Universidade de S\~ao Paulo, S\~ao Paulo - CEP 05314-970 SP, Brazil\\
$^2$Instituto de F\'isica - Universidade Federal do Paran\'a, Curitiba - CEP 81531-990 PR, Brazil\\
$^3$ Departamento de F\'isica - UNESP, Rio Claro - CEP 13506-900 SP, Brazil}

\date{\today} \widetext

\pacs{05.45.-a, 05.45.Pq, 05.40.Fb}

\begin{abstract}
The changeover from normal to super diffusion in time dependent 
billiards is explained analytically. The unlimited energy growth for an 
ensemble of bouncing particles in time dependent billiards is obtained by means 
of a two dimensional mapping of the first and second moments of the velocity 
distribution function. We prove that for low initial velocities the mean velocity 
of the ensemble grows with exponent $\sim1/2$ of the number of collisions with the 
border, therefore exhibiting normal diffusion. Eventually, this regime changes to a 
faster growth characterized by an exponent $\sim1$ corresponding to super diffusion. For 
larger initial velocities, the temporary symmetry in the diffusion of velocities explains 
an initial plateau of the average velocity.
\end{abstract}
\maketitle

As coined by Enrico Fermi \cite{Ref1} Fermi acceleration (FA) is a 
phenomenon where an ensemble of classical and non interacting particles 
acquires energy from repeated elastic collisions with a rigid and time 
varying boundary. It is typically observed in billiards 
\cite{Ref2,Ref3,Ref4} whose boundaries are moving in time 
\cite{Ref5,Ref6,Ref7,Ref7b,Ref7c}. If the motion of the boundary is 
random and the initial velocity is small enough \cite{Ref8}, the growth 
of the average velocity is proportional to $n^{1/2},$ with $n$ denoting 
the number of collisions. If the initial velocity is larger, a plateau 
of constant velocity is observed in a plot ${\overline{V}}~vs.~n$ which 
is explained from the symmetry of the velocity diffusion 
\cite{Ref9}. The symmetry warrants that part of the ensemble grows and 
part of it decreases in such a way the growing parcel cancels the 
portion decreasing. As soon as such symmetry is broken the constant 
regime is changed to a regime of growth. For deterministic oscillations 
of the border, the scenario is different. Breathing oscillations 
preserve the shape but not the area of the billiard. It is known that
the average velocity evolves in a sub-diffusive manner with a slope of 
the order of $1/6$ \cite{Ref10,Ref11}. For oscillations preserving the 
area but not the shape of the billiard there are two regimes of growth. 
For short time the diffusion of velocities is normal passing to super 
diffusion regime for large enough number of collisions \cite{Ref11b}. 
This changeover is, so far, not yet explained and our contribution in 
this letter is to fill up this gap in the theory. This is achieved by 
studying the momenta of the velocity distribution function, noticing
that the dynamical angular/time variables have an inhomogeneous 
distribution in phase space.

\begin{figure}[b]
\centerline{\includegraphics[width=0.28\textwidth]{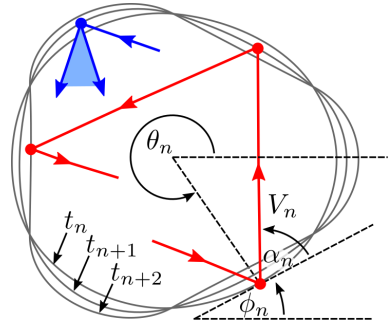}}
\caption{\label{fig1} (Color online) Three consecutive collisions of a 
particle in a deterministic time time dependent billiard (red 
trajectory). Illustration of the reflection angle range for a billiard 
with random motion in the boundary (blue). The relevant angles and 
velocity are shown for the $n$th collision.}
\end{figure}

The results presented in this letter are illustrated by a time-dependent 
oval-billiard \cite{Ref12} whose phase space is mixed when the boundary 
is static. The boundary of the billiard is written as 
$R_{b}(\theta,t)=1+\epsilon\left[1+a\cos(t)\right]\cos(p\theta)$ where
$R_{b}$ is the radius of the boundary in polar coordinates, $\theta$ is 
the polar angle, $\epsilon$ controls the circle deformation, $p>0$ is an 
integer number \cite{Ref13} given the shape of the boundary, $t$ is the 
time and $a$ is the amplitude of oscillation of the boundary. Figure 
\ref{fig1} shows a typical scenario of the boundary and three collisions 
illustrating the dynamics. 

\begin{figure}[ht]
\includegraphics[width=0.49\textwidth]{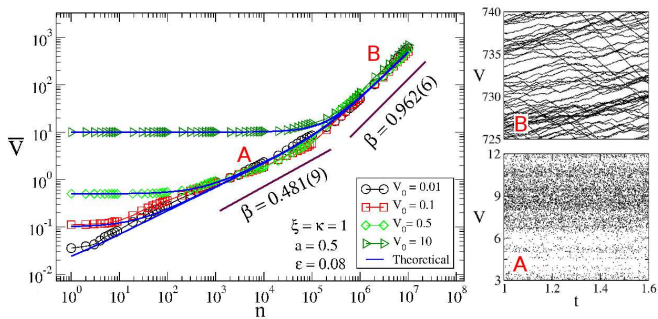}
\caption{\label{fig2} (Color online) Plot of $\overline{V}~vs.~n$ for 
different initial velocities, as labeled in the figure. Three 
different regimes are clear in the figure. For large initial 
velocities, a constant plateau dominates the dynamics for short $n$. 
After a first crossover, the average velocity starts to grow as a 
power law diffusing the velocity as a normal diffusion with slope of 
$0.481(9)$. Soon after, there is a second crossover where the normal 
diffusion is replaced by a super diffusion with slope $0.962(6)$. 
The right panels show portions of a single realization in the 
$(V,t)$-space where $A$ identifies normal diffusion and $B$ super 
diffusion. The parameters used are $\epsilon=0.08$, $a=0.5$ and $p=3$.}
\end{figure}

The dynamics of the particle is given in terms of a discrete, nonlinear 
and four dimensional mapping of the type 
$A:R^4\rightarrow R^4$ that transforms the dynamical 
variables at collision $n$ to their new values at collision $n+1$, 
$(\theta_{n+1},\alpha_{n+1},V_{n+1},t_{n+1}) = 
A(\theta_{n},\alpha_{n},V_{n},t_{n})$ where $V_n = |\vec{V}_n|$ denotes 
the magnitude of the velocity particle after collision $n$, and 
$\alpha_n$ corresponds the angle between the particle trajectory and the 
tangent line at the $n^{th}$ collision with the boundary at the polar 
angle $\theta_n$ and collision time $t_n$ (see fig. \ref{fig1}). Given 
that each particle moves along a straight line between collisions and 
with constant speed, the radial position of the particle is given by 
$R_p(t)=\sqrt{X^2(t)+Y^2(t)}$, where $X(t)$ and $Y(t)$ are the 
rectangular coordinates of the particle at the time $t$. The angular 
position $\theta_{n+1}$ is obtained by the numerical solution of 
$R_p(\theta_{n+1},t_{n+1})=R_b(\theta_{n+1},t_{n+1})$. The 
time at collision $n+1$ is given by $t_{n+1}=t_n+{{\sqrt{\Delta 
X^2+\Delta Y^2}} \over \vert\vec{V_n}\vert}$, where $\Delta 
X=X_p(\theta_{n+1},t_{n+1})-X(\theta_{n},t_{n})$ and $\Delta 
Y=Y_p(\theta_{n+1},t_{n+1})-Y(\theta_{n},t_{n})$. Because the 
referential frame of the boundary is non inertial, a change of 
referential must be made before the application of the momentum 
conservation law. The reflection laws at the instant of collision
are $\vec{V^{\prime}}_{n+1}\cdot\vec{T}_{n+1}=\xi\vec{V^{\prime}}_n\cdot 
\vec{T}_{n+1},\label{law1}$ and 
$\vec{V^{\prime}}_{n+1}\cdot\vec{N}_{n+1}=-\kappa\vec{V^{\prime}}_n\cdot 
\vec{N}_{n+1}$, where the unit tangent and normal vectors are 
$\vec{T}_{n+1}=\cos(\phi_{n+1})\hat{i}+\sin(\phi_{n+1})\hat{j}$,  
$\vec{N}_{n+1}=-\sin(\phi_{n+1})\hat{i}+\cos(\phi_{n+1})\hat{j}$ 
and $\xi,\kappa\in[0,1]$ are the restitution coefficients for the 
tangent and normal directions. The term $\vec{V^{\prime}}$ corresponds 
the velocity of the particle measured in the non-inertial reference 
frame. The tangential and normal components of the velocity after 
collision $n+1$ are
\begin{eqnarray}
\vec{V}_{n+1}\cdot\vec{T}_{n+1}&=&\xi\vec{V}_{n}\cdot\vec{T}_{n+1}+(1-\xi)\vec{V}_{b}\cdot\vec{T}_{n+1},\\
\vec{V}_{n+1}\cdot\vec{N}_{n+1}&=&-\kappa\vec{V}_{n}\cdot\vec{N}_{n+1}+(1+\kappa)\vec{V}_{b}\cdot\vec{N}_{n+1},
\end{eqnarray}
where $\vec{V}_{b}(t_{n+1})={dR_b(t)\over {dt}}\Big\vert_{t_{n+1}}
[\cos(\theta_{n+1})\widehat{i}+\sin(\theta_{n+1})\widehat{j}]$ is the 
velocity of the moving boundary at time $t_{n+1}$. The velocity of the 
particle after the collision $n+1$ is given by
${V}_{n+1}=\sqrt{(\vec{V}_{n+1}\cdot\vec{T}_{n+1}
)^2+(\vec{V}_{n+1}\cdot\vec{N}_{n+1})^2}$, while the angle 
$\alpha_{n+1}$ is written as
$\alpha_{n+1}=\arctan \left[{\vec{V}_{n+1}\cdot\vec{N}_{n+1} \over
\vec{V}_{n+1}\cdot\vec{T}_{n+1}} \right]$.

\begin{figure}[ht]
\includegraphics[width=0.4\textwidth]{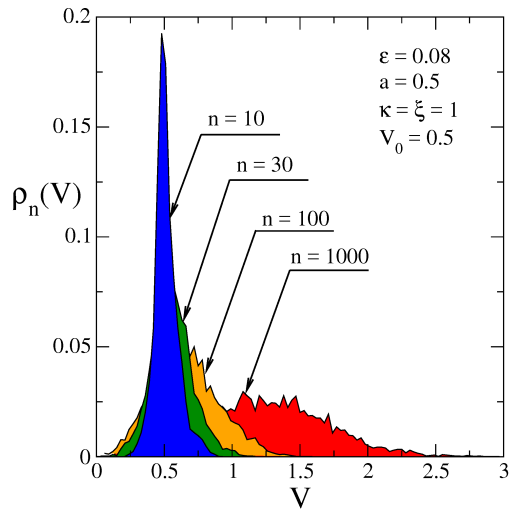}
\caption{\label{fig3} (Color online) Plot of the evolution of the 
velocity distribution function $\rho_n(V)$ for an ensemble of 
$5000$ particles after different number of collisions. The parameters used are 
$\epsilon=0.08$, $a=0.5$ and $p=3$.}
\end{figure}

Given an initial condition $(\theta_n,\alpha_n,V_n,t_n)$, the 
dynamical properties of the system can be obtained through application 
of the previous equations. We are interested in the behavior of the 
average velocity, obtained from two different kinds of average, namely 
$\overline V_n = {1\over 
M}\sum_{i=1}^{M}{1\over{n+1}}\sum_{j=0}^{n}|\vec{V}|_{i,j},$
where the first summation is made over an ensemble of different initial 
conditions randomly distributed in $t\in[0,2\pi]$, $\alpha\in[0,\pi]$ 
and $\theta\in[0,2\pi]$ for a fixed initial velocity while the second 
summation corresponds to an average made over the orbit, hence in time. 
We considered $M=5000$ and $n=10^7$ collisions. Figure \ref{fig2} 
shows the behavior of $\overline{V}~vs.~n$ for different initial 
velocities. Three different kinds of behavior can be seem from the 
figure. If the initial velocity is large, the curve of average velocity 
exhibits a constant plateau. The size of the plateau depends on the 
value of the initial velocity (see Ref. \cite{Ref9} for a discussion of 
such kind of behavior in a two dimensional mapping). A higher  initial 
velocity, leads to a longer plateau. A first crossover is observed 
changing the behavior of the constant regime to the regime of growth 
with a slope of growth typical of normal diffusion. A numerical fitting 
gives a slope $0.481(9)$. The regime of normal diffusion then reaches a 
second crossover passing to a faster regime of growth named as super 
diffusion with slope $0.962(6)$. The panels on the right hand side of 
fig. \ref{fig2}, give the plot of a single realization in the 
$(V,t)$-space where $A$ corresponds to normal diffusion and $B$ to 
super diffusion. We emphasize when the perturbation on the boundary is 
random, the dynamics in the $(V,t)$-space is similar to what is observed 
in $A$ with a constant slope of growth about $1/2$ therefore 
characterized by normal diffusion. When the restitution coefficients 
$\xi,\kappa\ne1$, inelastic collisions occur leading to a different 
scenario \cite{Ref14}, where the energy growth is interrupted by the 
violation of Liouville's theorem and attractors emerge in the phase 
space.

The explanation of the initial plateau is related to the behavior of 
the velocity distribution function \cite{Ref14}. For 
an initial velocity larger than the maximum moving wall velocity, say 
$V_{max}=a\epsilon$, the following is observed: (1) Part of the 
ensemble of particles acquires energy leading to such portion of the 
ensemble to grow energy; (2) However, another part of the ensemble 
leads to decreasing of velocity. Each parcel cancels each other 
producing the constant plateau. The decrease however is limited to 
the lower bound of the velocity, in this case, null velocity. When 
the particles reach the velocity lower bound, the symmetry of the 
velocity diffusion is broken leading to a crossover 
between the constant regime and the growth regime. Figure 3 shows the 
evolution of the velocity distribution function $\rho_n(V)$, for an 
ensemble of particles as the number of collisions is increased.

\begin{figure}[t]
\centerline{\includegraphics[width=0.5\textwidth]{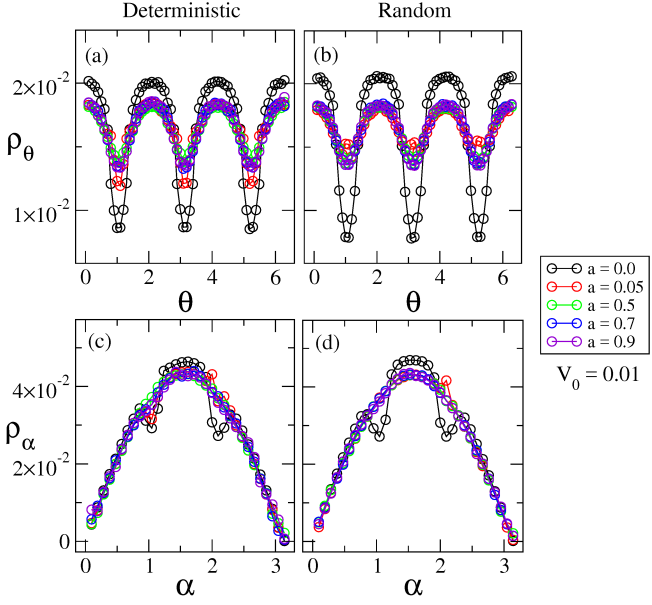}}
\caption{\label{fig4} (Color online) Plot of the numerical distribution 
functions $\rho_\theta(\theta)=\int\tilde F(\theta,\alpha)d\alpha$ 
and $\rho_\alpha(\alpha)=\int\tilde F(\theta,\alpha)d\theta$ for 
deterministic and non-deterministic (random) billiards at various 
amplitude of oscillations. The parameters used are 
$\epsilon=0.08$, $a=0.5$ and $p=3$.}
\end{figure}
\begin{figure}[t]
\centerline{\includegraphics[width=0.5\textwidth]{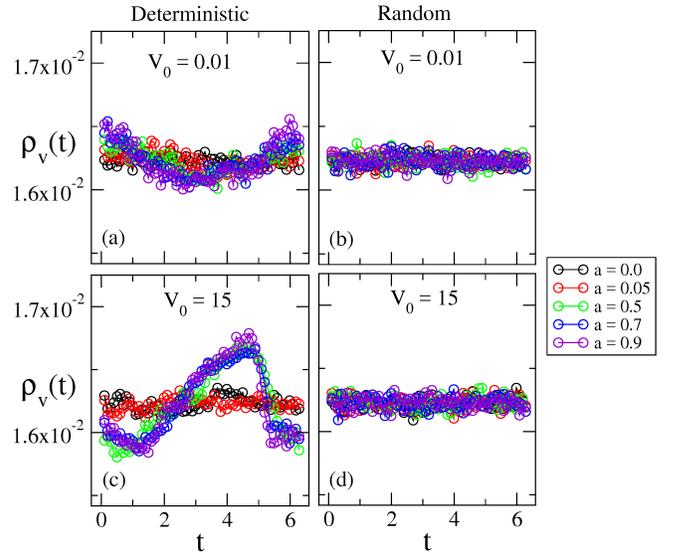}}
\caption{\label{fig5} Plot of the numerical collision time distribution functions 
$\rho_V(t)$ for deterministic and non-deterministic (random) billiards 
at various amplitude of oscillations and two different initial 
velocities, as shown in the figure. The parameters used are 
$\epsilon=0.08$, $a=0.5$ and $p=3$.}
\end{figure}

The velocity of a particle after collision with the wall 
$\tilde V$, can be written in the form
\begin{equation}\label{eq.collision_rule}
 \tilde V(\alpha,\theta,t,V) = V + \zeta(\alpha,\theta, t, V),
\end{equation}
where $V$ is the velocity just before the collision. From this, the 
mean velocity of an ensemble of particles just after the $n^{th}$ 
collision takes the form
\begin{equation}\label{eq.V_n+1}
 \overline{V}_{n+1} = \overline{V}_n + \delta\overline{V}_n,
\end{equation}
where 
\begin{eqnarray}\label{eq.deltaV}
\overline{V}_n &= \int_0^\infty \!\! \int_0^{2\pi} \!\! \int_0^{2\pi}
 \!\! \int_0^\pi& V\mathcal{F}_n(\alpha,\theta,t,V) d\alpha d\theta dt dV,\\
 \delta\overline{V}_n &= \int_0^\infty \!\! \int_0^{2\pi} \!\! \int_0^{2\pi}
 \!\! \int_0^\pi& \zeta(\alpha,\theta, t, V)\mathcal{F}_n(\alpha,\theta,t,V)\times \nonumber \\
 && d\alpha d\theta dt dV, 
\end{eqnarray}
and $\mathcal{F}_n(\alpha,\theta,t,V)$, is the phase space distribution 
function {just before collision $n$}. In the case of interest the 
distribution function can be factored in the form
\begin{equation}\label{eq.Fn_factor3}
 \mathcal{F}_n(\alpha,\theta,t,V) = F(\theta,\alpha)\rho_V(t)\rho_n(V),
\end{equation}
where $F(\alpha,\theta)$ is the $\alpha-\theta$ distribution, 
$\rho_{V}(t)$ is the collision time distribution and $\rho_n(V)$ is the 
velocity distribution function. The $F(\alpha,\theta)$ distribution is mainly 
determined by the billiard geometry and is only weakly modified by 
the wall oscillation, consequently it can be regarded as independent of 
the index $n$, the collision time distribution $\rho_V(t)$ depends on the 
velocity but not on the index $n$ and the velocity distribution function $\rho_n(V)$ 
depends on both the velocity and the index. To understand the evolution of the 
mean velocity of the ensemble we do not require to determine the evolution of the 
global distribution function. Inserting eq. (\ref{eq.Fn_factor3}) in eq. (\ref{eq.deltaV}), 
and defining the partial mean
\begin{eqnarray}
 U(V) &= \int_0^{2\pi} \!\! \int_0^{2\pi} \!\! \int_0^\pi & \zeta(\alpha,\theta, t, V) F(\theta,\alpha)\rho_V(t) d\alpha d\theta dt, \quad
\end{eqnarray}
we obtain a compact expression for the change in the mean velocity
\begin{eqnarray}\label{eq.deltaVn}
\delta\overline{V}_n &= \int_0^\infty& \rho_n(V)U(V)dV,
\end{eqnarray}
where the partial mean $U(V)$ can be expanded around the mean velocity 
$\overline{V}_n$
\begin{equation}\label{eq.U(V)_approx}
U(V)\approx U(\overline{V}_n) + U'(\overline{V}_n)(V-\overline{V}_n) + 
{1\over{2}} U''(\overline{V}_n)(V-\overline{V}_n)^2+...
\end{equation}

This approximation becomes poor as we move far from the distribution mean. 
However, the velocity distribution $\rho_n(V)$ drops for large and 
small values of $V$, and the integrand in eq. (\ref{eq.deltaVn}) vanishes 
where the Taylor expansion is not accurate. Inserting eq. (\ref{eq.U(V)_approx}) 
in eq. (\ref{eq.deltaVn}) and replacing in eq. (\ref{eq.V_n+1})
we obtain a second order approximation of the $n+1$ mean velocity
\begin{equation}
\overline{V}_{n+1} = \overline{V}_n + U(\overline{V}_n) + 
{1\over 2}U''(\overline{V}_n)(\overline{V^2_n}-\overline{V_n}^2),
\end{equation}
which depends on the quadratic mean of the velocity at
collision $n$. An equation for the evolution of the quadratic mean is 
also needed. For the quadratic velocity after collision $\tilde V^2$,
the collision rule can be also written in the form
\begin{equation}\label{eq.Vprime2_generic}
 \tilde V^2(\alpha,\theta,t,V) = V^2 + \psi(\alpha,\theta,t, V),
\end{equation}
where $V$ is again the velocity before collision. Reproducing the same
arguments used for the mean velocity we obtain
\begin{equation}
\overline{V^2}_{n+1} = \overline{V^2}_n + W(\overline{V}_n) + 
{1\over2}W''(\overline{V}_n)(\overline{V^2_n}-\overline{V_n}^2),
\end{equation}
where
\begin{eqnarray}\label{eq.W(V)}
W(V)\!\! &= \int_0^{2\pi} \!\! \int_0^{2\pi} \!\! \int_0^\pi& \!\! \psi(\alpha,\theta, t, V)
F(\theta,\alpha)\rho_V(t) d\alpha d\theta dt. \qquad
\end{eqnarray}

As will be shown later the inhomogeneous nature of $F(\theta, \alpha)$ is a fundamental
aspect of super diffusion. It can be related to the presence of low period saddles in
the static billiard \cite{Ref15}, where the collisions occur more often \cite{Ref16}, leading to an 
increase in the distribution value. Figure \ref{fig4} we show line-integrated profiles 
of $F(\alpha,\theta)$ for both deterministic and random oscillations of the billiard boundary.

In the limit of high velocities or small amplitudes of the wall motion 
it can be shown that $\zeta(\alpha,\theta,t,V) \approx {\psi(\alpha,\theta,t,V)}/{2V},$
which leads to
\begin{equation}
 U(V) \approx {W(V)}/{2V}.
\end{equation}

This results in an approximated form for the two-dimensional
mapping of the velocity mean and the quadratic velocity mean are
\begin{eqnarray}
  \overline{V}_{n+1} &=& \overline{V}_n + {1\over2}{W_n}{\overline{V^2_n}}/{\overline{V_n}^3} + {1\over2}\left({1\over2}\overline{V_n}W''_n -W'_n\right) \nonumber \\
  && \times\left({\overline{V_n^2}}/{\overline{V_n}^2}-1\right),\nonumber\\
 \overline{V^2}_{n+1} &=& \overline{V^2}_n + W_n + {1\over2}W''_n(\overline{V^2_n}-\overline{V_n}^2),\label{eq.velocity_mapping}
\end{eqnarray}
where $W_n = W(\overline{V_n}), W'_n=W'(\overline{V_n})$ and 
$W''_n=W''(\overline{V_n})$. This mapping is general and its
its behavior depends on the particular form of the function $W(V)$ for 
the system under consideration. For instance, for the time-dependent 
oval billiard one can show
\begin{eqnarray}
 \psi (\alpha,\theta,t) &=& 4 (a\epsilon)^2\cos^2(p\theta)\sin^2(t)+ \\ \nonumber
  && - 4a\epsilon V\cos(p\theta)\sin(\alpha)\sin(t),
\end{eqnarray}
and the collision time distribution can be approximated by (see fig.~\ref{fig5}(a,c))
\begin{equation}
 \rho_V(t) = {1\over2\pi}\left[1-a\epsilon\chi(V)\sin(t)\right],
\end{equation}
where $\chi(V)$ is a slowly changing function of $V$ that vanishes for 
$V=0$ and saturates at $\chi^*$ for large $V$. This distribution develops due
to the correlation between subsequent collisions for higher velocities for
which the time between collisions is small compared to the wall oscillation
period. Expectedly, the harmonic part is removed when the wall oscillations are
random (see fig.~\ref{fig5}(b,d)).

Inserting $\psi(\alpha,\theta,t)$ and $\rho_V(t)$ in eq. (\ref{eq.W(V)}) and defining 
the following constants
\begin{eqnarray}
\eta_1 &= (a\epsilon)^2\int_0^\pi\int_0^{2\pi}& F(\alpha,\theta)\sin(\alpha)\cos(p\theta)d\theta d\alpha,\\
 \eta_2 &= (a\epsilon)^2\int_0^\pi\int_0^{2\pi}& F(\alpha,\theta)\cos^2(p\theta)d\theta d\alpha,
\end{eqnarray}
we obtain
\begin{equation}
 W(V) = 2\eta_2 + 2\eta_1\chi(V)V.
\end{equation}
which inserted in the mappings for the mean velocity and the quadratic 
mean (eq. (\ref{eq.velocity_mapping})), results in two coupled difference 
equations
\begin{eqnarray}
 \overline{V_{n+1}}-\overline{V_n} &=& \eta_1\chi_n + \eta_2{\overline{V_n^2}}/{\overline{V_n}^3},\label{eq.difV}\\
 \overline{V^2_{n+1}}-\overline{V^2_n} &=& 2\eta_2 + 2\eta_1\chi_n\overline{V_n},\label{eq.difV2}
\end{eqnarray}
where $\chi_n = \chi(\overline{V_n})$, and it was used that $\chi(V)$ 
changes slowly with the velocity. More specifically the system satisfies 
$\chi'(\overline{V_n})\overline{V_n}<<\chi(\overline{V_n})$. These 
coupled difference equations can be solved asymptotically for small and
large $n$ to give us a picture of the different diffusion regimes 
exhibited by the system during its evolution.

If the ensemble of particles begins with small velocities, i.e. on the order of $a\epsilon$,
we can use $\chi_n\rightarrow 0$, and the system can be integrated by taking the
continuous limit $\overline{f}_n-\overline{f}_{n-1}\approx df(n)/dn$,
which results in
\begin{eqnarray}
 \overline{V^2_s}(n) = V_0^2 + 2\eta_2 n,
\end{eqnarray}
where the sub-index $s$ indicates small $n$. This solution can be replaced 
in the mean velocity difference to obtain an approximated solution for the
mean velocity valid for small velocities and few collisions
\begin{eqnarray}
 \overline{V_s}(n) = \left(V_0^2 + 2\eta_2 n\right)^{1/2}.
 \label{Vs}
\end{eqnarray}
Notice this solution emerged from assuming an homogeneous phase 
distribution, i.e. $\chi_n\rightarrow 0$, which is also appropriate when the 
collision phase with the wall is random. However, for the deterministic 
case, as the ensemble of velocities grows, the phase distribution 
becomes inhomogeneous and $\chi_n$ saturates to $\chi^*$. As this
occurs the velocities disperse in the velocity space, and provided that
$\overline{V_n^2}$ and $\overline{V_n}^2$ are of the same order we have
that $\overline{V_n^2}/\overline{V_n}^3\rightarrow 0$. From this, the
mean velocity satisfies approximately
\begin{equation}
 \overline{V_{n+1}}-\overline{V_n} \approx \eta_1\chi^*,
\end{equation}
which can be integrated to obtain the evolution rule for the 
high-velocities regime
\begin{equation}
\overline{V_l}(n) = V_0 + \eta_1\chi^*n.
\label{Vh}
\end{equation}
Here, the sub-index $l$ indicates large $n$. For regular values of
$\eta_1, \eta_2$ and $\chi^*$, the difference $V_l(n)-V_0$ is small compared
to $V_s(n)-V_0$ for small $n$, while the opposite happens for large $n$.
Consequently, we can combine eq. (\ref{Vs}) and eq. (\ref{Vh}) to obtain a single
approximate solution valid for all stages of the ensemble evolution
\begin{equation}\label{eq.combined}
 \overline{V}(n) = (V_0^2 + 2\eta_{2}n)^{1/2}+\eta_{1}\chi^*n.
\end{equation}

An interesting feature of this solution is that $\eta_1$ vanishes if 
$F(\alpha,\theta)$ is homogeneous, so that, a deterministic billiard 
without low period saddles in phase space will only exhibit normal 
diffusion because of the uniform distribution of particles in the phase 
space. The combined solution eq. (\ref{eq.combined}), corresponds to the 
continuous lines in fig. \ref{fig2} in excellent agreement with the 
numerical simulations for all the ensembles considered.

As a short summary, in this letter we have shown that an inhomogeneous 
particle distribution function on the phase space of the static 
billiard leads to the development of anomalous diffusion regimes in
time-dependent situations, and for the particular case of oval billiards,
explains the transitions from normal to super diffusion. The presented treatment,
however, is sufficiently general to study other anomalous diffusion regimes, 
diverse billiard shapes and more general mappings.

\acknowledgments
MH thanks to CAPES for financial support. DC, ILC
and EDL thanks to CNPq (433671/2016-5, 300632/2010-0 and 303707/2015-1) and ILC 
and EDL thanks to São Paulo Research Foundation (FAPESP) (2011/19296-1 and 2017/14414-2) 
for their financial support.


\begin{thebibliography}{0}
\bibitem{Ref8} As small we refer as to velocities from the order of 
the maximum moving boundary velocity.
\end{thebibliography}
\end{document}